\documentclass[letter,traditabstract]{aa} %
\usepackage{graphicx}
\usepackage{natbib}
\usepackage{amssymb}
\usepackage{amsmath}
\usepackage{url}
\usepackage[section]{placeins}
\usepackage{multirow}
\usepackage{txfonts}
\begin{document}
   \title{Dissecting the Moth: Discovery of an off-centered ring in the HD~61005 debris disk with high-resolution imaging\thanks{Based on observations collected at the European Southern Observatory, Chile, ESO program 0184.C-0567(E)}}

\author{E. Buenzli\inst{\ref{inst1}} 
\and C. Thalmann \inst{\ref{inst2}}
\and A. Vigan \inst{\ref{inst4}}
\and A. Boccaletti \inst{\ref{inst3}}
\and G. Chauvin \inst{\ref{inst5}}
\and J.C. Augereau \inst{\ref{inst5}}
\and M.R. Meyer \inst{\ref{inst1}}
\and F. M\'enard \inst{\ref{inst5}}
\and S. Desidera \inst{\ref{inst6}}
\and S. Messina \inst{\ref{inst10}}
\and T. Henning \inst{\ref{inst2}}
\and J. Carson \inst{\ref{inst11},}\inst{\ref{inst2}}
\and G. Montagnier \inst{\ref{inst7}}
\and J.L. Beuzit \inst{\ref{inst5}}
\and M. Bonavita \inst{\ref{inst9}}
\and A. Eggenberger \inst{\ref{inst5}}
\and A.M. Lagrange \inst{\ref{inst5}}
\and D. Mesa \inst{\ref{inst6}}
\and D. Mouillet \inst{\ref{inst5}}
\and S. P. Quanz \inst{\ref{inst1}}
}

\institute{
Institute for Astronomy, ETH Zurich, 8093 Zurich, Switzerland\label{inst1}\\
\email{ebuenzli@astro.phys.ethz.ch}
\and
Max Planck Institute for Astronomy, Heidelberg, Germany\label{inst2}
\and
Laboratoire d'Astrophysique de Marseille, UMR 6110, CNRS, Universit\'e de Provence, 13388 Marseille, France\label{inst4}
\and 
LESIA, Observatoire de Paris-Meudon, 92195 Meudon, France \label{inst3}
\and
Laboratoire d'Astrophysique de Grenoble, UMR 5571, CNRS, Universit\'e Joseph Fourier, 38041 Grenoble, France\label{inst5}
\and
INAF - Osservatorio Astronomico di Padova, Padova, Italy \label{inst6}
\and 
INAF - Osservatorio Astrofisico di Catania, Italy \label{inst10}
\and 
College of Charleston, Department of Physics \& Astronomy, Charleston, South Carolina, USA\label{inst11}
\and
European Southern Observatory: Casilla 19001, Santiago 19, Chile\label{inst7}
\and
University of Toronto, Toronto, Canada \label{inst9}
}

\date{received 22 September 2010 / accepted 09 November 2010} \titlerunning{The off-centered debris ring around HD~61005}




\abstract
{The debris disk known as ``The Moth'' is named after its
unusually asymmetric surface brightness distribution.  It is located around
the $\sim$90\,Myr old G8V star HD~61005 at 34.5\,pc and has previously been imaged by the HST at 1.1 and 0.6\,$\mu$m. Polarimetric observations suggested that the circumstellar material consists of two distinct components, a nearly edge-on disk or ring, and a swept-back feature, the result of interaction with the interstellar medium. We resolve both components at unprecedented resolution with VLT/NACO $H$-band imaging. Using optimized angular differential imaging techniques to remove the light of the star, we reveal the disk component as a distinct narrow ring at inclination $i=84.3 \pm 1.0\degr$. We determine a semi-major axis of $a=61.25 \pm 0.85$ AU and an eccentricity of $e=0.045\pm0.015$, assuming that periastron is located along the apparent disk major axis. Therefore, the ring center is offset from the star by at least $2.75 \pm 0.85$ AU. The offset, together with a relatively steep inner rim, could indicate a planetary companion that perturbs the remnant planetesimal belt. From our imaging data we set upper mass limits for companions that exclude any object above the deuterium-burning limit for separations down to $0\farcs3$. The ring shows a strong brightness asymmetry along both the major and minor axis. A brighter front side could indicate forward-scattering grains, while the brightness difference between the NE and SW components can be only partly explained by the ring center offset, suggesting additional density enhancements on one side of the ring. The swept-back component appears as two streamers originating near the NE and SW edges of the debris ring. 
}


\maketitle


\section{Introduction} 

Dust in planetary systems is most likely produced by collisions of planetesimals that are frequently arranged in a ring-like structure \citep{wyatt08}. 
These debris disks are therefore thought to be brighter analogs of our solar system's Kuiper belt or asteroid belt. Previous studies found no correlation between the presence of known massive planets and infrared dust emission from debris disks \citep{moro-martin07,bryden09, kospal09}, but several systems are known to host both (e.g. HR~8799, HD~69830). Scattered light imaging of debris disks has revealed numerous structures thought to be shaped by planets. Warps in the $\beta$ Pic debris disks \citep{mouillet97,augereau01} are caused by a directly confirmed $9\pm3\,M_\mathrm{J}$ planet \citep{lagrange09,lagrange10,quanz10}. Fomalhaut hosts a debris ring with a sharp inner edge and an offset between ring center and star \citep{kalas05b}, for which dynamical models suggest the presence of a planet \citep{quillen06}. A planetary candidate was indeed imaged \citep{kalas08}.
Other larger scale asymmetric structures, e.g. around HD~32297 \citep{kalas05a}, are thought to result from interaction with the ambient interstellar medium (ISM), such as movement through a dense interstellar cloud \citep{debes09}. Alternatively, they might be perturbed by a nearby star \citep[e.g. HD~15115,][]{kalas07}.

The source HD~61005 (G8V, $V=8.22$, $H=6.58$, $d=34.5\pm1.1$\,pc) was first discovered to host a debris disk by the Spitzer/FEPS program \citep[formation and evolution of planetary systems,][]{meyer06}. It has the largest 24\,$\mu$m infrared excess with regard to the photosphere of any star observed in FEPS \citep[$\sim$110\%,][]{meyer08}. The star's age was estimated to be $90\pm40$\,Myr \citep{hines07}. The disk was resolved with HST/NICMOS coronagraphic imaging at 1.1\,$\mu$m \citep{hines07} that revealed asymmetric circumstellar material with two wing-shaped edges, that is the reason for in the nickname ``the Moth''. The disk was also resolved at 0.6\,$\mu$m with HST/ACS imaging and polarimetry \citep{maness09}. The polarization suggests two distinct components, a nearly edge-on disk or ring and a swept-back component that interacts with the ISM. Both papers focused on the properties and origin of the swept-back component. In this work, we present high-contrast ground-based imaging with unprecedented angular resolution. We discover and characterize a distinct asymmetric ring in the inner disk component and discuss the probability of a planetary companion.

   \begin{figure}[tb]
   \centering
   \resizebox{0.49\textwidth}{!}
            {\includegraphics{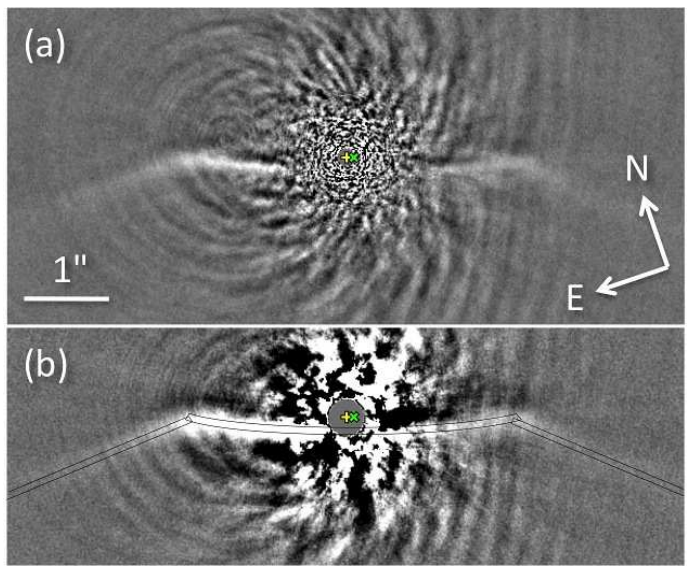}}
      \caption{High-contrast NACO $H$-band images of HD~61005, \textbf{(a)} reduced with LOCI, 
      \textbf{(b)} reduced with ADI. The slits used for photometry are overlaid. The curved slit traces the maximum surface brightness of the lower ring arc, while the rectangular boxes enclose the streamers. In both images the scaling is linear, and 1\arcsec corresponds to a projected separation of 34.5\,AU. The arc-like structures in the background are artifacts of the observation and reduction techniques, and are asymmetric because the field rotation center was offset from the star. The region with insufficient field rotation is masked out. The plus marks the position of the star, the cross the ring center.}
         \label{fig:reduc}
   \end{figure}

\section{Observations and data reduction}

We observed HD~61005 on February 17, 2010 with the NACO instrument \citep{rousset03,lenzen03} at the VLT. The observations were obtained in the framework of the NaCo Large Program Collaboration for Giant Planet Imaging (ESO program 184.C0567). The images were taken in the $H$-band (1.65 $\mu$m) in pupil-tracking mode \citep{kasper09} to allow for angular differential imaging \citep[ADI,][]{marois06}. The field of view was $14\arcsec\times14\arcsec$ and the plate scale 13.25\,mas/pixel. We performed the disk observations without a coronagraph, and used the cube-mode of NACO to take 12 data cubes. Each cube consisted of 117 saturated exposures of 1.7927\,s, yielding a total integration time of 41.95\,min. The saturation radius was $\sim$0\farcs15. A total of 112$^\circ$ of field rotation was captured while the pupil remained fixed. Before and after the saturated observations we took unsaturated images with a neutral density filter to measure the photometry for the central star. The adaptive optics system provided a point spread function (PSF) with a full-width at half-maximum (FWHM) of 60\,mas with $\sim$0\farcs8 natural seeing in $H$-band (22\% Strehl ratio). 

The data were flat-fielded, bad-pixel corrected, and centered on the star by manually determining the center for the middle frame and aligning the others through cross-correlation. We removed 3 bad-quality frames and averaged the remaining images in groups of three for a total of 467 frames. We then used LOCI \citep[locally optimized combination of images,][]{lafreniere07} and customized ADI to subtract the stellar PSF to search for point sources and extended non-circular structures. 

In ADI, each image is divided into annuli of 2 FWHM width. For each frame and each annulus, a frame where the field object has rotated by 2 FWHM is subtracted to remove the stellar halo. Additionally, the resulting image is subtracted by a back-rotated version of itself. Finally, all images are derotated and median combined. In LOCI, each annulus is further divided into segments, and for each segment an optimized PSF is constructed from a linear combination of sufficiently rotated frames. A minimum rotation of 0.75\,FWHM is optimal for point source detection and has led to several detections around other targets \citep{marois08,thalmann09,lafreniere10}. To reveal the extended nebulosity around LkCa~15, \citet{thalmann10} used a much larger minimum separation of 3\,FWHM. For the nearly edge-on and therefore very narrow debris disk around HD~61005, we obtain an optimal result for a minimum separation of 1\,FWHM, but using large optimization areas of 10000 PSF footprints to lessen the self-subtraction of the disk. We also reduced the data with LOCI with a separation criterion of 0.75\,FWHM and small optimization segments of 300 PSF footprints to set hard detection limits on companions.

Additionally we attempted a classical PSF subtraction using a reference star, which is detailed in Appendix \ref{sec:psf}.

\section{Results}

The NACO $H$-band images obtained by reduction with LOCI and ADI are shown in Fig.~\ref{fig:reduc}.
The circumstellar material is resolved to an off-centered, nearly edge-on debris ring with a clear inner gap and two narrow streamers originating at the NE and SW edges of the ring. A strong brightness asymmetry is seen between the NE and SW side and between the lower and upper arc of the ring. The inner gap has not been previously resolved by HST, where only the direction of the polarization vectors hinted at a disk-like component separate from the extended material that  interacts with the ISM. 

LOCI provides the cleanest view of the ring geometry with respect to the background because it effectively removes the stellar PSF while bringing out sharp brightness gradients. The negative areas near the ring result from oversubtraction of the rotated disk signal embedded in the subtracted PSF constructed by LOCI. In particular, the ring's inner hole is enhanced. However, tests with artificial flat disks showed that while self-subtraction can depress the central regions, the resulting spurious gradients are shallow and different from the steep gradients obtained from the edge of a ring. Because of significant variable flux loss, photometry is unreliable in the LOCI image. In the ADI reduction, the self-subtraction is deterministic and can be accounted for, while the stellar PSF is subtracted adequately enough to allow photometric measurements. 
In the image produced by reference PSF subtraction (Fig.~\ref{fig:psf}) the stellar PSF is not effectively removed. The image is unsuitable for a quantitative analysis, but it confirms the streamers and the strong brightness asymmetry, and also suggests the presence of a gap on the SW side. 

\begin{figure}[tb]
   \centering
      \resizebox{0.85\hsize}{!}
            {\includegraphics{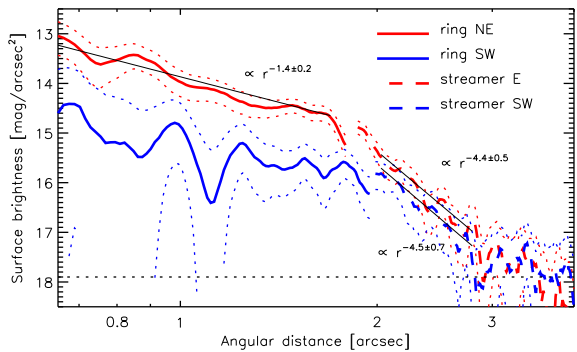}}
   \caption{$H$-band surface brightness of the brighter ring arc and the streamers measured from the ADI image and corrected for self-subtraction. Dotted lines indicate the error. The dotted dark line is the 1\,$\sigma$ sky background. Solid dark lines are power-law fits with the obtained slope indicated. The transition regions between ring and streamers are not fitted.}
      \label{fig:sb}%
      \vspace*{-5mm}
\end{figure}

\subsection{Surface brightness of ring and streamers}

The surface brightness of the ring and streamers (Fig. \ref{fig:sb}) is obtained from the ADI image. We measure the mean intensity of the bright ring arc as a function of angular separation from the star in a curved slit of 5 pixels width (see Fig. 1b) following the maximum brightness determined in Sect. \ref{sec:offset}. For the streamers the slit is rectangular and of the same width. We calculate the mean intensity in the intersection of the slit with annuli of 9 pixels width. To estimate the self-subtraction because of ADI, we apply our ADI method to a model ring (Sect. \ref{sec:offset}). The measured flux loss in our slit is $24\pm5\%$, where the error includes variations with radius and between the two sides. For the true ring this value might differ by a few percent. For lack of a good model for the swept material, we apply the same correction factor for the streamers, though the systematic error is likely larger. The dominant source of error are subtraction residuals, which we measure as the dispersion in wider slits rotated by $\pm$45\degr{}. To obtain absolute photometry we use the observations of HD~61005 taken in the neutral density filter as reference. The NE arc is about 1 mag/arcsec$^2$ or a factor of 2--3 brighter than the SW arc, consistent with the factor of 2 brightness asymmetry seen by HST at shorter wavelengths. The surface brightness of the inner 1\farcs1 of the SW arc is of the same order as the residuals, making a power-law fit unreliable. 

The two streamers are tilted by an angle of $\sim$23\degr{} with respect to the ring's semi-major axis. We detect material out to a projected distance of $\sim$140\,AU (4\arcsec) from the star. Beyond 2.8\arcsec{} the S/N ratio is too low to perform a meaningful power-law fit. Closer, the power-law slopes agree with those by \citet{maness09} within errors. We do not detect the fainter, more homogeneously distributed swept material seen by HST because such structures are subtracted by ADI. However, the streamers are also seen in the reference PSF subtracted image and thus are the most visible component of the swept material. They may represent the limb-brightened edges of the total scattering material.

\subsection{Ring geometry and center offset}\label{sec:offset}

We convolve the LOCI image with the PSF and measure the ring's inclination and position angle by ellipse fitting through points of maximum intensity in selected regions. Assuming that the ring is intrinsically circular, the fit yields an inclination of $84.3^\circ$ and position angle $70.3^\circ$, with systematic errors of $\sim$1$^\circ$. The position angle agrees well with the position angle of the disk-component determined by HST, while we find the inclination to be $\sim$4\degr{} closer to edge-on. To determine the separation of the ring ansae, we create inclined ring annuli and find for each side the ring with maximum mean intensity around the ansa. This fit yields a radius of $61.25\pm0.85$ AU, and a ring center offset from the star by $2.75\pm0.85$ AU toward SW along the apparent disk major axis. The radial extension of the ring agrees with the location of the power-law break seen by HST at 0.6\,$\mu$m.

To assess the effect of our PSF subtraction method and determine additional ring parameters, we create synthetic scattered light images with the GRaTer code for optically thin disks \citep{augereau99}. The models are inserted into the data of our PSF reference star (cf.\ Appendix \ref{sec:psf}) and the LOCI algorithm is applied to allow direct comparison with the observations. We let the ring be intrinsically elliptical with the periastron located along the apparent disk major axis, because our data cannot constrain an offset along the minor axis. The radial structure is described by a smooth combination of two power laws, rising to the ring peak density position and fading with distance from the star. Scattering is represented by a Henyey-Greenstein (HG) phase function with an asymmetry parameter $g$.

\begin{figure}[tb]
   \centering
      \resizebox{0.85\hsize}{!}
	  {\includegraphics{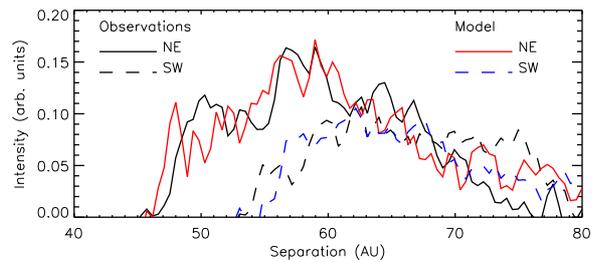}}
	   \caption{Comparison of a radial cut averaged over 5 pixels through the midplane for the observation reduced with LOCI and the model disk implanted into the reference star and reduced in the same way. The intensity of the SW-side of the model is scaled down by a factor 1.3.}
      \label{fig:radcut}
	  \vspace*{-5mm}
\end{figure}

We compare the model constructed from the parameters derived above with the observation. An inclination of $i=84.3\pm1.0^\circ$ is also a good match for an intrinsically elliptical ring. While the center offset leads to a small increase (factor of $\sim$1.2) in surface brightness on the NE side with respect to the SW side, models that are identical on the two sides except for the offset still underestimate the extent of the enhancement by a factor of $\sim$1.3. That the brightness asymmetry is visible after all reduction methods and in the HST 0.6\,$\mu$m image suggests a physical asymmetry in the density or grain properties. Because asymmetric dust models that include the ISM interaction go beyond the scope of this Letter, we focus on validating the ring geometry. 

We compare a radial cut along the midplane averaged over 5 pixels (Fig. \ref{fig:radcut}), artificially lowering the model intensity on the faint side by a factor of 1.3. Indeed, models with an offset $o = a \cdot e$ of $2.75 \pm 0.85$\,AU, where $a$ is the semi-major axis for the peak density ($ 61.25\pm 0.85$\,AU) and $e$ the eccentricity ($0.045 \pm 0.015$), still provide a decent match after reduction with LOCI. Models without offset are worse particularly out to $\sim$63 AU for each ring side because the shift in peak intensity is missing. Therefore the offset does not appear to be an artifact of the data reduction.

Model comparison suggests an inner surface density power-law slope of $\sim$7, but a fit is difficult because reduction artifacts differ for models and observations. The outer slope (fixed to $-4$) is uncertain because we do not model the ISM interaction. In any case, the inner rim appears to be significantly steeper than the outer rim.
From the brightness asymmetry between the upper and lower arc we estimate the asymmetry parameter to $|g| \sim$0.3. This value is uncertain because the weak arc is strongly contaminated by reduction residuals. Additionally, the HG phase function is a simplistic model for scattering in debris disks. A positive $g$-value, assuming that the brighter side is the front, would indicate forward scattering grains. This may not always be the case \citep[see e.g.][]{min10}.

\subsection{Limits on companions to HD~61005}

In our full image (Fig. \ref{fig:bgobjects}) we detect six point sources at $r > 3\arcsec$. These are seen in the \citet{hines07} data as well. Astrometric tests show that their relative proper motion is consistent with all objects being background sources (cf.\ App.\ \ref{sec:astrometry}).

\begin{figure}[tb]
   \centering
      \resizebox{0.85\hsize}{!}
            {\includegraphics{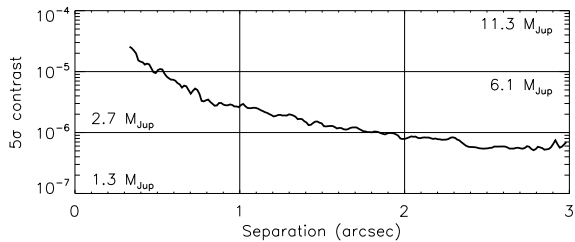}}
   \caption{Contrast for companions around HD~61005 detectable at the 5$\sigma$ level. The numbers indicate the mass limit at the horizontal lines for an age of 90 Myr, based on the COND models by \citet{baraffe03}. One arcsec corresponds to a projected separation of 34.5\,AU.}
      \label{fig:masslimit}%
\end{figure}

In the LOCI image reduced with the smaller minimum rotation we search for closer companions. After convolving the resulting image with an aperture of 5 pixels diameter, we calculate the noise level at a given separation as the standard deviation in a concentric annulus. To determine the flux loss from partial self-subtraction we implant artifical sources in the raw data. The measured contrast curve is corrected for this flux loss to yield the final $5\sigma$ detectable constrast curve (Fig.~\ref{fig:masslimit}). We translate the contrast to a mass limit using the COND evolutionary models by \citet{baraffe03}. We assume an age of 90\,Myr. We do not detect any companion candidates, but are able to set limits well below the deuterium burning limit.

\section{Discussion}

The high-resolution image enables us to distinguish the actual debris ring from the material that appears to be streaming away from the system. The results reveal a ring center offset of $\sim$3\,AU and an additional brightness asymmetry suggesting density variations. The eccentricity of the debris ring could be shaped by gravitational interaction with a companion on an eccentric orbit. A similar system is Fomalhaut with a belt eccentricity of 0.11 \citep{kalas05b}. Mass constraints were discussed for Fomalhaut b by \citet{chiang09}. A rough adaptation of their result to HD 61005 shows that a planet below our detection limit of $\sim$3-4\,$M_\mathrm{J}$ located beyond $\sim$40~AU at maximum angular separation could perturb the ring. Because of the high inclination a planet of higher mass and lower semi-major axis could also hide within the residuals at smaller projected separation from the star.

Models of the spectral energy distribution by \citet{hillenbrand08} suggested that the debris required multiple temperature components. This could be fitted by either an extended debris model ($R_\mathrm{inner} < 10$ AU and $R_\mathrm{outer} > $40 AU) or more likely an inner warm ring and an outer cool ring, which could coincide with the dust detected in the scattered light images. 

To explain the structure of the interacting material, \citet{maness09} explored several scenarios, and proposed that a low-density cloud is perturbing grain orbits because of ram pressure. The streamers would be barely bound, sub-micron sized grains on highly eccentric orbits, consistent with the observed blue color and the brightness profile. Their model currently cannot reproduce the sharpness of the streamers, but the observed geometry of the parent body ring might help improve the models to validate this theory. These might then answer whether the ring offset could also be caused by the ISM interaction rather than a planet. Obtaining colors of the ring through high-resolution imaging at other wavelengths could indicate if a grain size difference exists between parent body ring and swept material.

As a solar-type star, and with a debris ring at a radius not much larger than that of the Kuiper belt, the HD~61005 system provides an interesting comparison to models of the young solar system. \citet{booth09} calculate the infrared excess of the solar system as a function of time based on strong assumptions consistent with the Nice model. At $\sim$90\,Myr the calculated 70\,$\mu$m excess ratio $F_{70}$/$F_{70\star}$ is about four times lower than that observed for HD~61005 \citep{hillenbrand08}. Indeed, HD~61005 is one of the most luminous debris disks known. Its other obvious unusual feature is the morphology of the swept component. Perhaps further modeling will provide a causal connection between its high observed dust generation rate and this apparent interaction with the ISM. It is also an interesting target for future deeper searches for planetary mass companions as well as remnant gas that could be associated with the debris.

\bibliographystyle{aa}
\bibliography{15799bib}

\begin{acknowledgements}
We thank David Lafreni{\`e}re for providing the source code for his LOCI algorithm. We thank J. Alcala, E. Brugaletta, E. Covino and A. Lanzafame for discussions on the stellar properties and H.M. Schmid for comments. EB acknowledges funding from the Swiss National Science Foundation (SNSF). AE is supported by a fellowship for advanced researchers from the SNSF (grant PA00P2\_126150/1).
\end{acknowledgements}

\Online

\begin{appendix}

\section{Subtraction of stellar halo with a PSF reference star} \label{sec:psf}

\begin{figure}[hbt]
   \centering
   \resizebox{0.49\textwidth}{!}
            {\includegraphics{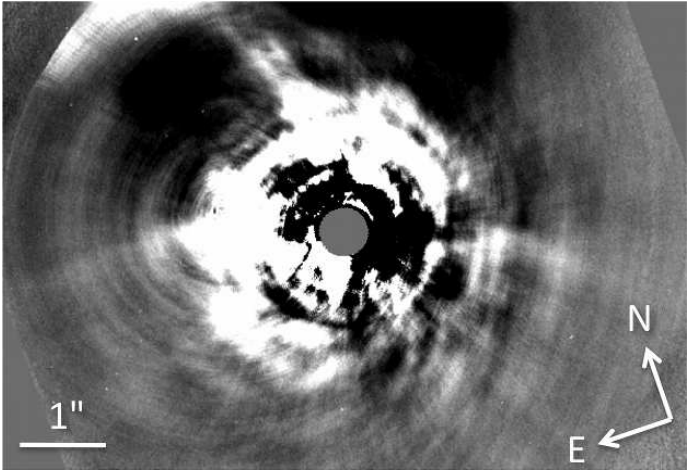}}
      \caption{Reference PSF subtracted image of HD~61005. }
         \label{fig:psf}
   \end{figure}

To check that the revealed structures are not artifacts of the ADI methods, a classical PSF subtraction was applied to HD~61005 using the unresolved star TYC-7188-571-1, observed 3 hours after HD~61005 in the same observing mode. A total of 400 frames matching the same parallactic angle variation for HD~61005 and the reference star were considered. After shift-and-add, a scaling factor was derived from the ratio of the azimuthal average of the HD~61005 median image to the azimuthal average of the recentered median image of the PSF. The median of the PSF sequence was then subtracted from all individual frames of HD~61005. The resulting cube was derotated and collapsed to obtain the final PSF-subtracted image (Fig. \ref{fig:psf}). Additional azimuthal and low-pass filtering was applied to improve the disk detection.

\section{Astrometry of background sources} \label{sec:astrometry}

In addition to our VLT data, we used the HST/NICMOS observations of
Hines et al. (2007) (program-10527) obtained in November 20, 2005 and
June 18, 2006 to identify the status of the six faint sources
detected in our VLT/NaCo field (marked as fs-1, 2, 3, 4, 5 and 6 in
Fig. A.1). The relative positions recorded at different epochs can be
compared to the expected evolution of the position measured at the
first epoch under the assumption that the sources are either
stationary background objects or comoving companions. For the range of
explored semi-major axes, any orbital motion can be considered to be of
lower order compared with the primary proper and parallactic
motions. Considering a proper motion of
$(\mu_\alpha,\mu_\delta)=(-56.09\pm0.70,74.53\pm0.65)$~mas/yr and a
parallax of $\pi = 28.95\pm0.92$~mas for HD\,61005 as well as the
relative positions of all faint sources at each epoch (see Table \ref{tab:new_binaries}),
a $\chi2$ probability test of $2\times N_{epochs}$ degrees of freedom
(corresponding to the measurements: separations in the $\Delta\alpha$
and $\Delta\delta$ directions for the number $N_{epochs}$ of epochs)
was applied. None of the six sources are comoving with
HD~61005 with a probability higher than $99.99\%$. They are found to
be background stationary objects with a probability higher than $60\%$.
We can therefore fully exclude the possibility that these sources are
physically bound companions of HD~61005. \citet{maness09} had already determined that four sources visible in their image were background sources. Two of these correspond to fs-3 and fs-4, and we here confirm their result. Their other two sources are outside of our field of view. We therefore provide a new result for the four sources fs-1, fs-2, fs-5 and fs-6.

\begin{table}[t]
\caption{Relative positions of the faint sources 1 to 6 (Fig. \ref{fig:bgobjects}). A conservative astrometric error of 1 pixel has been considered
 for the relative position measurements obtained with HST/NICMOS and
 VLT/NaCo observations (i.e 75.8~mas and 13.25~mas).}
\label{tab:new_binaries}
\begin{tabular}[\columnwidth]{llll}     
\hline\hline
\noalign{\smallskip}
Name   &  UT Date   &     $\Delta$RA      &    $\Delta$DEC      \\
      &            & (mas)        & (mas)    \\
\noalign{\smallskip}\hline\noalign{\smallskip}

fs-1   &   2006-06-18  &   -1929    &     2918\\
      &   2010-02-17  &   -2189 &      3080\\
fs-2   &   2005-11-20  &   -4672 &        4669\\
      &   2010-02-17  &   -4449 &        4441\\
fs-3   &   2005-11-20  &   -2326 &        9321\\
      &   2010-02-17  &   -2068 &        9034 \\
fs-4   &   2005-11-20  &   -8487 &         -15\\
      &   2010-02-17  &   -8222 &        -370\\
fs-5   &   2006-06-18  &   -1586 &       -6848\\
      &   2010-02-17  &   -1358 &       -7123\\
fs-6   &   2006-06-18  &   -1148 &       -7671\\
      &   2010-02-17  &   -1336 &       -8024\\

\noalign{\smallskip}\hline\noalign{\smallskip}
\noalign{\smallskip}\hline                 \end{tabular}
\end{table}

\begin{figure}[bt]
   \centering
      \resizebox{\hsize}{!}
            {\includegraphics{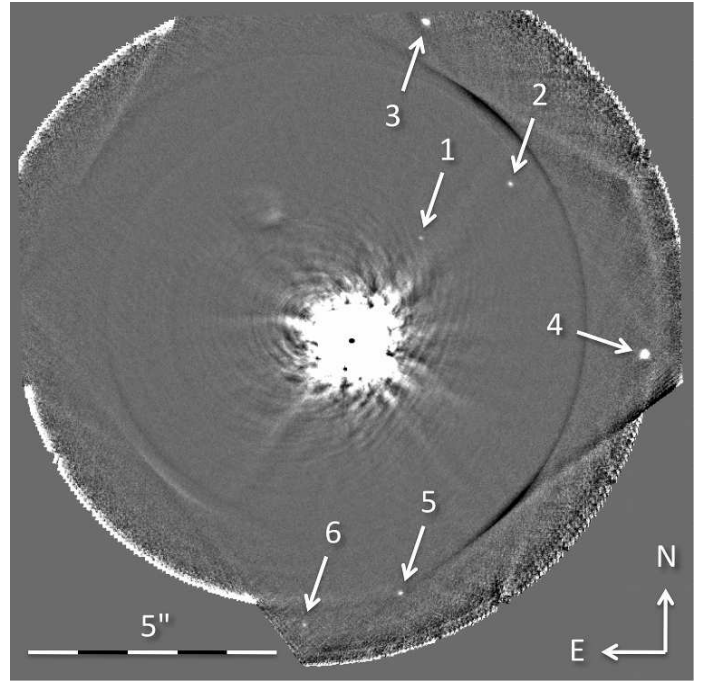}}
   \caption{Full field of view of our NACO $H-$band data reduced by derotating, adding and spatially filtering. Six background sources are identified.}\label{fig:bgobjects}
\end{figure}

\end{appendix}

\end{document}